\documentclass[american,letterpaper,12pt]{article}
\usepackage{babel}
\usepackage[latin1]{inputenc}
\usepackage{graphicx}
\usepackage{graphics}
\usepackage{amsmath}
\usepackage[centerlast,small]{}
\usepackage{color}
\usepackage{lmodern}
\usepackage[T1]{fontenc}
\usepackage{textcomp}
\title{ Predictions on the execution of a CNOT gate of the reverse of the arrow of time  }

\author{G. Morales \thanks{giselle.mora17@gmail.com}\\ M. \'Avila \\
Centro Universitario UAEM Valle de Chalco, UAEMex\\
Mar\'{\i}a Isabel, Valle de Chalco\\
Estado de M\'exico, CP 56615. \\
F. Soberanes\thanks{fabian\_ sm@tesch.edu.mx}\\Tecnologico de Estudios
Superiores de Chalco\\Tlapala, Chalco\\ Estado de M\'exico, CP 56641.}

\begin{document}
\maketitle

\begin{abstract}
Recently it has been pointed out that an outstanding
application of an IBM quantum computer is to reverse
the arrow of time [Lesovik et al.
Sci. Rep. {\bf 9}, 1 (2019)]. The issue of the consequences of the reversal of the arrow of time on the execution
of a CNOT gate by a quantum device is addressed. 
It is predicted that if the CNOT gate is executed towards the future then this was previously executed in the past.
The above might confirm the paradigm that an event in the present time influences another event in the past time \cite{schmidt}. Such a physical phenomena is called
retrocausality.
\end{abstract}
%\begin{document}

\maketitle

% \noindent PACS: 03.65.-w; 03.65.Aa; 03.67.Bg; 03.67.-a

 \newpage
A considerable amount of interest has been paid to the
crucial observation done by Lesovik {\it et al} \cite{lesovik} in the sense
that the arrow of time can be inverted in an IBM quantum computer. Undoubtedly the above is one of the most outstanding application
of a quantum computer. It is worth to observe also that
the result of Ref. \cite{lesovik} sheds light on the enigma of the correlation
between time-symmetric fundamental
laws of Physics and irreversibility of the world \cite{lloyd}-\cite{wigner}. Furthermore, the above
pave the way for
a further revolutionary technological applications of a quantum computer.
However, so far it has not been explored the issue of the consequences on the execution of a basic quantum gate 
by a quantum device of the reverse of the arrow of time. In the present work it is addressed such an issue. 
In order to proceed, we consider such a quantum device as the 
diamond quantum computer. It is worth to observe that a diamond quantum computer is a very promising technology
with outstanding capabilities \cite{lopez}-\cite{lopez2}.
Then it is verified first, that this quantum computer executes the CNOT gate towards the future.
After that, it is inverted the time towards the past and studied the consequence of the above on the execution
of the CNOT gate. It is predicted that such a quantum gate was previously executed in the past. 
At this stage it result important to observe that the above result might confirm the old paradigm of retrocausality.

A quantum computer must work by executing of a primitive operations named quantum gates.
  A CNOT gate \cite{nielsen} is one of the the most primitive operation that can be performed by a quantum
  computer. At this stage it is worth to observe that a quantum computer architecture is often described by a Schr\"oedinger equation
  with a suitable Hamiltonian \cite{majer}-\cite{yu}.
  In particular, the execution of several different
  quantum protocols in a diamond quantum computer can be performed
  with the use of a Schroedinger equation with a RF
  based Hamiltonian \cite{lopez}-\cite{lopez2}.
    Within the above approach we first verify that a diamond quantum computer
    can execute a CNOT gate towards the future in the sense of the Second Law of
    Thermodynamics \cite{coveney}. Then, it is inverted the time and studied the consequences of the above on the execution of the CNOT gate.
  The present work is organized as follows: it is first introduced
the Hamiltonian describing a diamond quantum computer.
Then it is proved that a diamond quantum computer executes
a CNOT gate from the present time towards the past time.
Finally a series of conclusions are given.

The corresponding Schr\"oedinger equation associated the
diamond quantum computer is \cite{lopez}

\begin{eqnarray}
 i \hbar \frac{\partial}{\partial t} \mid \psi (t) \rangle &  =  &
 \Bigl\{-\hbar \left(w_1  S_{1}^{Z} + w_2  S_{2}^{Z} \right) -  \frac{J}{\hbar} \left( S_{1}^{Z}  S_{2}^{Z} \right) +
  \nonumber \\ & &  \frac{\Omega}{2} \Bigl[ \left( e^{i\theta_1}   S_{1}^{-} + e^{-i\theta_1}   S_{1}^{+} \right) +  \nonumber \\ & &
  \left( e^{i\theta_2}   S_{2}^{-} +
  e^{-i\theta_2}   S_{2}^{+} \right)
   \Bigr]
  \Bigr\}\mid \psi (t) \rangle ,
\end{eqnarray}

\noindent where $\hbar=1$ is Planck's constant,
$w_1$ ( $w_2$) are Larmor's frequencies, $S_{1}^{Z}$  ($S_{2}^{Z}$)
are the $z$-components of the spin of qubit 1 (qubit 2),  $S_{1}^{-}$
($S_{2}^{-}$) are the lowering spin operator of the
qubit 1 (qubit 2), $S_{1}^{+}$ ($S_{2}^{+}$)
are the raising spin operator of the qubit 1 (qubit 2),
$J$ is a coupling constant\footnote{If $J>0$ the diamond quantum computer is in
an antiferromagnetic regime but if $J<0$ this is in a ferromagnetic regime \cite{bonechi}.},
$\Omega$ is Rabi's frequency, and $\theta_i=w_i t + \varphi_i$ ($i=1,2$) being $\varphi_i$
($i=1,2$) an arbitrary phases.

By observing that a general two-qubit state can be written as

\begin{eqnarray}
\mid \psi(t)\rangle =C_0(t) \mid 00 \rangle +
C_1(t) \mid 01 \rangle+C_2(t) \mid 10 \rangle+C_3(t) \mid 11 \rangle,
\end{eqnarray}

\noindent where

\begin{eqnarray}
  \mid C_0(t)\mid^2 + \mid C_1(t)\mid^2 +\mid C_2(t)\mid^2+\mid C_3(t)\mid^2=1,
\end{eqnarray}

then Eq. (1) can also be written as

\begin{eqnarray}
 i	\frac {d C_0(t)}{dt} = {\left[ - \left( w_1 + w_2 \right) - J \right] C_0(t) + \left( \frac{\Omega}{2} \cdot e^{-i \theta_2} \right) C_1(t) + \left( \frac{\Omega}{2} \cdot e^{-i \theta_1} \right)C_2(t)}, \nonumber\\
i\frac{d C_1(t)}{d t} = {\left( \frac{\Omega}{2} \cdot e^{i \theta_2} \right)C_0(t) + \left[ -  ( w_1 - w_2) + J \right] C_1(t) + \left( \frac{\Omega}{2} \cdot e^{-i \theta_1} \right) C_3(t)},
\nonumber \\
\\
i\frac	{d C_2(t)}{d t} = {\left( \frac{\Omega}{2} \cdot e^{i \theta_1} \right)C_0(t) + \left[ -  (- w_1 + w_2) + J \right] C_2(t) + \left( \frac{\Omega}{2} \cdot e^{-i \theta_2} \right) C_3(t)},
\nonumber  \\
i	\frac{d C_3(t)}{d t} = {\left( \frac{\Omega}{2} \cdot e^{i \theta_1} \right)C_1(t) + \left( \frac{\Omega}{2} \cdot e^{i \theta_2} \right)C_2(t) + \left[ -  (-w_1-w_2) -J \right] C_3(t)}. \nonumber
\end{eqnarray}

\vskip0.5cm

\noindent {\it - Execution of a CNOT gate towards the future}

\vskip0.3cm

In this case the arrow of time is taken according to the Second Law of Thermodynamics
i.e. from the past towards the future.

Given the initial conditions

\begin{eqnarray}
C_0(t=0)=0, \nonumber \\
C_1(t=0)=0, \nonumber \\
\\
C_2(t=0)=1, \nonumber \\
C_3(t=0)=0. \nonumber
\end{eqnarray}

\noindent It is said that the diamond quantum computer
described by Eq. (1) executes a CNOT gate in
a scale of time $t=T$ ($T>0$) if

\begin{eqnarray}
\mid C_0(t=T)\mid^2 =0, \nonumber \\
\mid C_1(t=T)\mid^2 =0, \nonumber \\
\\
\mid C_2(t=T)\mid^2 =0, \nonumber \\
\mid C_3(t=T)\mid^2 =1. \nonumber
\end{eqnarray}

In order to verify the conservation
of probability of Eq. (3) towards the future ($t>0$), it is first solved numerically
Schr\"oedinger equation. The results
are shown in Figure 1.
Together with the above, it is also verified that a diamond quantum computer executes
the CNOT gate of Eq. (6) towards the future. The above is shown in Figures 2, 3, 4, and 5.
The employed values for the constants are
$J=0.0015$, $\Omega=0.01$, $\varphi_1=\pi/2$, $\varphi_2=\pi/4$,
$w_1=0.2$, and $w_2=0.0015$.

\vskip0.5cm

\noindent {\it - Consequences of the reverse of the arrow of time on the execution of a CNOT gate}

\vskip0.3cm

It is said that the diamond quantum computer executes
the CNOT gate towards the past
if given the initial conditions of Eq. (5) then
\footnote{It is clear that negative times do not exist in nature.
We employ mathematically this notation as a way
of expressing the reverse of the arrow of time.}

\begin{eqnarray}
\mid C_0(t=-T)\mid^2 =0, \nonumber \\
\mid C_1(t=-T)\mid^2 =0, \nonumber \\
\\
\mid C_2(t=-T)\mid^2 =0, \nonumber \\
\mid C_3(t=-T)\mid^2 =1. \nonumber
\end{eqnarray}

By solving numerically Schr\"oedinger equation of Eq. (4)
for a negative times,
it is verified that the conservation of probability of Eq. (3)
was also preserved in the past.
The respective results are shown in Figure 1.
It is also shown that
the diamond quantum computer executed the CNOT gate of Eq. (7)
in the past. The above can be seen from Figures 2, 3, 4, and 5.
The employed values for the constants are
$J=0.0015$, $\Omega=0.01$, $\varphi_1=\pi/2$, $\varphi_2=\pi/4$,
$w_1=0.2$, and $w_2=0.0015$.

\vskip0.2cm

To be one of the most promising quantum technologies
for a quantum computer, it has been considered a diamond quantum computer as the
basis of our study. It has been checked first that such a computer
executes a CNOT gate with the arrow of time taken conventionally
according to the Second Law of Thermodynamics.
After that it is inverted the arrow of time and shown that the CNOT gate was executed in the past providing this was executed 
towards the future.
From present findings it is predicted that if a CNOT gate is executed towards the future then this was previously executed in the past. 
Such a result might confirm the physical phenomenon of retrocausality.

\vskip0.5cm

%These lines were commented in the \verb|IEEEtran.cls|:
%\begin{verbatim}
% IEEE uses Times font, so we'll default to times.
% These three commands make up the entire
% times.sty package.
\renewcommand{\sfdefault}{phv}
\renewcommand{\rmdefault}{ptm}
\renewcommand{\ttdefault}{pcr}
%\end{verbatim}

%\begin{figure}[htp]
%\centering {\includegraphics[height=2in,width=2in]{fig_03.JPG}}
%{\includegraphics[height=2in,width=2in]{fig_02.JPG}}
%\caption*{\hskip0.40cm (1a) \hskip4.0cm (1b)} \caption{(a) One and
%two qubits environment. (b) Many qubits environment.}
%\end{figure}

\begin{figure}[htp]
\centering {\includegraphics[height=2in,width=2in]{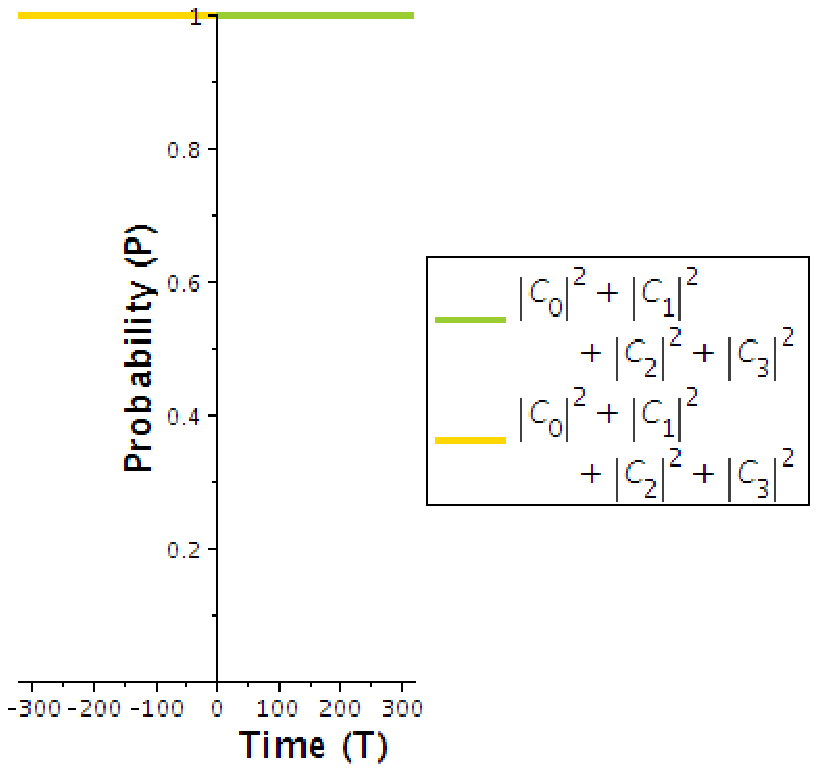}}
\caption{Conservation of the probability of Eq. (3) for both cases: Towards
the future ($t>0$) and Towards the past ($t<0$).}
\end{figure}

\begin{figure}[htp]
\centering {\includegraphics[height=2in,width=2in]{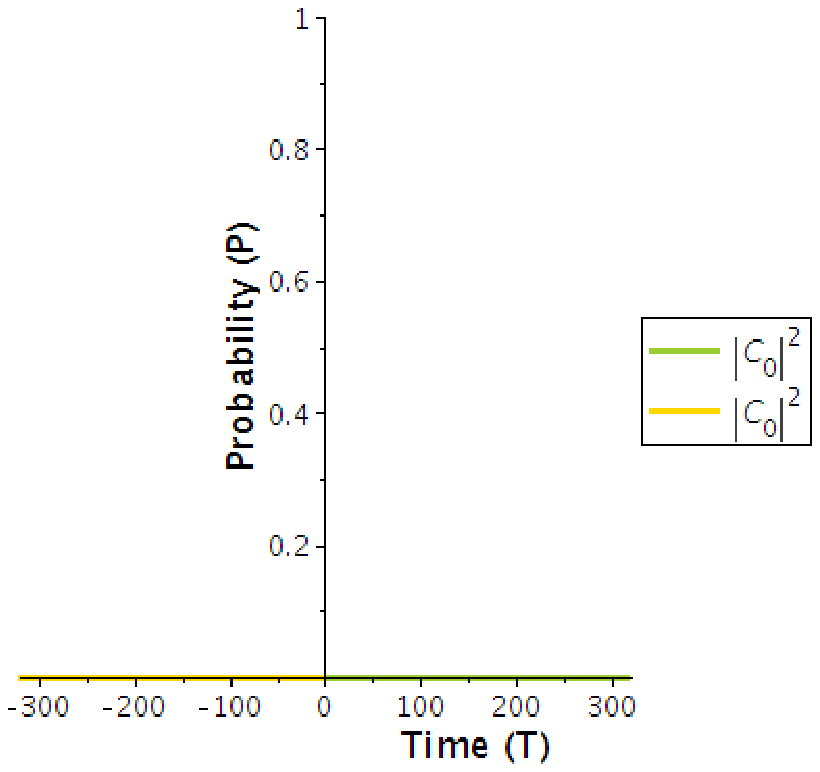}}
\caption{The coefficient $C_0(t)$ satisfies
the conditions of a CNOT gate of Eqs. (6) and (7)
for both cases towards the future ($t>0$)
and towards the past ($t<0$).}
\end{figure}

\begin{figure}[htp]
\centering {\includegraphics[height=2in,width=2in]{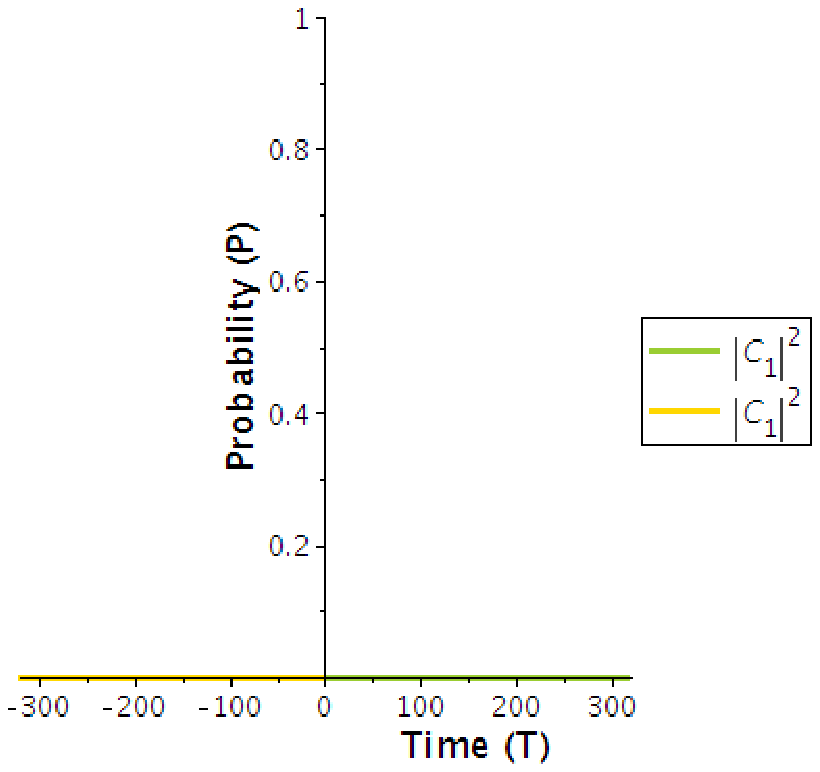}}
\caption{The coefficient $C_1(t)$
satisfies
the conditions of a CNOT gate of Eqs. (6) and (7)
for both cases towards the future ($t>0$)
and towards the past ($t<0$).}
\end{figure}

\begin{figure}[htp]
\centering {\includegraphics[height=2in,width=2in]{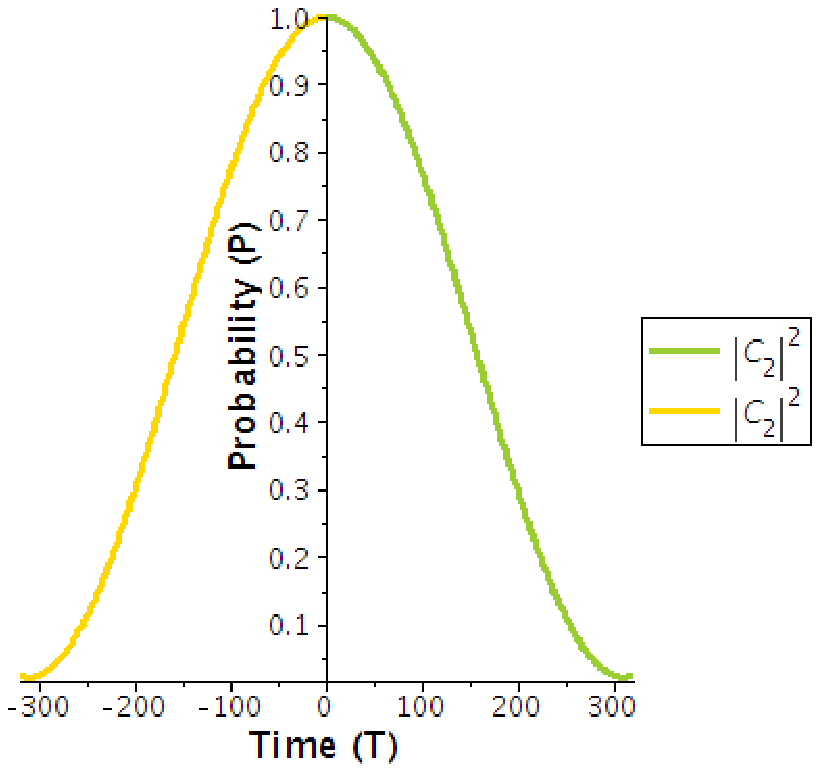}}
\caption{The coefficient $C_2(t)$ satisfies
the conditions of a CNOT gate
of Eqs. (6) and (7)  for both cases towards the future ($t>0$)
and towards the past ($t<0$).}
\end{figure}

\begin{figure}[htp]
\centering {\includegraphics[height=2in,width=2in]{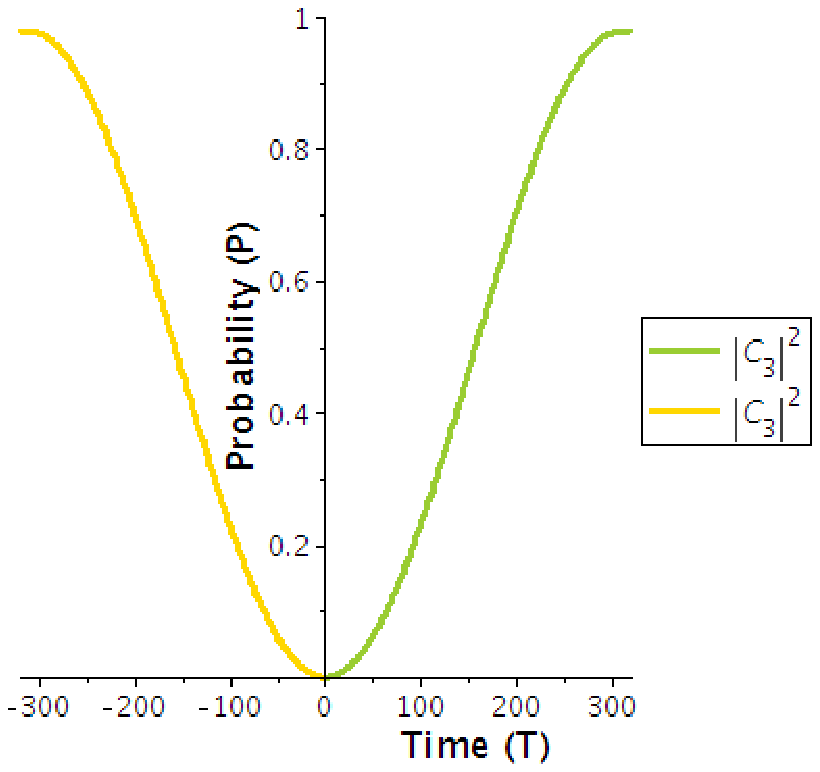}}
\caption{The coefficient $C_3(t)$ satisfies
the conditions of a CNOT gate
of Eqs. (6) and (7)  for both cases towards the future ($t>0$)
and towards the past ($t<0$).}
\end{figure}

%\begin{figure}[htp]
%\centering {\includegraphics[height=2in,width=2in]{Molecula_2.jpg}}
%\caption{Benzene (C$_6$H_6$_15$)molecule with six environmental
%qubits {E$_i$,\,\,\,\,\,\,$i=1,2,..,6$} and the information system
%composed by one qubit.}
%\end{figure}

%\begin{figure}[htp]
%\centering {\includegraphics[height=2in,width=2in]{mol_a.jpg}}
%\caption{Heptadecane(C$_{17}$H$_{36}$) molecule with 17
%environmental qubits and the information system composed by one
%qubit $\textnormal{q}_{\textnormal{0}}$. }
%\end{figure}

%Heptadecane(C$_{17}$H$_{36$}) molecule with 17 environmental qubits
%and the information system composed by one qubit

\newpage

\vskip1.5cm

%\begin{table}[!hbt]
%\begin{center}
%\begin{tabular}{|l|l|l|r|}
%\hline
%&I. Single bound &II. Double bound & III. Triple bound\\
%\hline
%$^{13}$C-H 2& 246.5 Hz & 277 Hz & \\
%\hline
%$^{13}$C-$^{13}$C & 173.8 Hz & 194 Hz & 250.2 Hz\\
%\hline
%\end{tabular}
%\caption{Values of the different environmental Zeeman couplings
%\cite{hart} employed in the calculations of the decoherence times of
%Eq. (27).}
%\end{center}
%\end{table}

\section*{
Acknowledgments} We thank SNI-Conacyt grant.

%\section{Figures}

%\begin{figure}
%\centering
%\scalebox{0.65}{ \input{fig1.ps} }
%\includegraphics[totalheight=1.20in]{fig1.ps}
%\includegraphics[totalheight=0.8\textheight,viewport=50 260 400 %1000,clip]{erptsqfit}
%\caption{Quantum computation by measuring two-state
%particles on a lattice. Before the measurements
%the qubits are in the cluster state $|\Phi>_{\mathcal C}$
%of (2). Circles $\odot$ symbolize measurements
%of $\sigma_z$, vertical arrows are measurements of
%$\sigma_z$, while tilded arrows refer to measurements
%in the $x-y$ plane.}
%\label{fig:FIGURE 1}
%\end{figure}

%\bibliographystyle{unsrt}
%\bibliography{refs}

\begin{thebibliography}{0}
\bibitem{lesovik} G. B. Lesovik, I. A. Sadovskyy, M. V. Suslov, A. V. Lebedev, and V. M. Vinokur
Sci. Rep. {\bf 9}, 1 (2019).
\bibitem{schmidt} H. Schmidt, Foundations of Physics {\bf 8}, 463 (1978).
\bibitem{lloyd} S. Lloyd, {\it On the spontaneous generation of complexity
in the universe in Complexity and the arrow of time}, ed. C. H. Lineweaver,
P. C. W. Davies, and M. Ruse (Cambridge University Press, 2013).
\bibitem{wigner} E. Wigner, Nachr. Ges. Wiss. G\"otingen, Math. Phys., K1
{\bf 1932}, 546 (1932).
\bibitem{nielsen} M. A. Nielsen and I. L. Chuang,
{\it Quantum Computation and Quantum Information}, Cambridge University Press (2000).
\bibitem{majer} J. Majer, I. M. Chow, I. M. Gambetta et al., Nature {\bf 449},
443 (2007).
\bibitem{lopez} G. Lopez, J. Lizarraga, J. of Mod. Phys. {\bf 11}, 1123 (2020).
\bibitem{lopez2} G. Lopez, G. LOpez, J. of Mod. Phys. {\bf 5} 55 (2014).
\bibitem{blais} A. Blais, R. -Sh. Huang,  A. Wallraff, et al., Phys. Rev. A {\bf 75}, 032329 (2007).
\bibitem{yu} Y. Yu, S. Han, et al., Science {\bf 296}, 889 (2002).
\bibitem{coveney} P. Coveney, R. Highfield, {\it The arrow of time}, Ed.
Ballantine Books (1992).
\bibitem{bonechi} F. Bonechi, E. Caleghini, R. Giachetti, E. Sorace,
M. Tarlini, J. of Phys. A {\bf 25} L939 (1992).
\end{thebibliography}
\end{document}